\documentclass[a4paper,fleqn]{cas-dc}    

\usepackage[numbers]{natbib}
\usepackage{booktabs}

\usepackage{float} 
\usepackage{placeins} 
\usepackage[version=4]{mhchem}
\usepackage{ulem}
\usepackage[switch]{lineno}

\ExplSyntaxOn
\cs_gset:Npn \__first_footerline: {}
\cs_gset:Npn \__second_footerline: {}
\ExplSyntaxOff

\makeatletter
\AtBeginDocument{%
  \thispagestyle{empty} 
}
\makeatother

\makeatletter
\providecommand{\doi}[1]{}          
\providecommand{\url}[1]{}          
\makeatother

\begin{document}
\let\WriteBookmarks\relax
\def\floatpagepagefraction{1}
\def\textpagefraction{.001}

\shorttitle{}    

\shortauthors{}  

\title [mode = title]{Tunable Magnetic and Topological Phases in EuMnXBi$_2$ \\(X=Mn, Fe, Co, Zn) Pnictides}  



%

\author[1]{Deep Sagar}

\affiliation[1]{organization={Indian Institute of Technology},
            addressline={School of Physical Sciences}, 
            city={Mandi},
            postcode={175005}, 
            state={Himachal Pradesh},
            country={India}}

\author[2]{Abhishek Sharma}

\affiliation[2]{organization={Indian Institute of Technology},
            addressline={School of Computing and Electrical Engineering}, 
            city={Mandi},
            postcode={175005}, 
            state={Himachal Pradesh},
            country={India}}

\author[1]{Arti Kashyap}

\cormark[1]

\ead{arti@iitmandi.ac.in}

\cortext[1]{Arti Kashyap}

\makeatletter
\@ifundefined{printorcid}{}{\let\printorcid\relax}
\makeatother
\makeatletter
\renewcommand\section{\@startsection{section}{1}{\z@}%
  {1.5\baselineskip plus .3\baselineskip minus .2\baselineskip}
  {1.5\baselineskip}
  {\normalfont\large}}                                             
  
\renewcommand\subsection{\@startsection{subsection}{2}{\z@}%
  {1.5\baselineskip \@plus 0.2\baselineskip \@minus 0.1\baselineskip}
  {1.5\baselineskip}
  {\normalfont\normalsize}}                                               
\makeatother


\begin{abstract}
 We present a comprehensive density functional theory (DFT) study of the electronic, magnetic, and topological properties of the layered pnictides EuMnXBi$_2$ (X = Mn, Fe, Co, Zn), focusing in particular on the relatively unexplored Bi-based member of the EuMn$_2$X$_2$ family. Unlike the well-studied As-, Sb-, and P-based analogues, we show that EuMn$_2$Bi$_2$ stabilizes in a C-type antiferromagnetic ground state with a narrow-gap semiconducting character. Inclusion of spin–orbit coupling (SOC) drives a transition from this trivial antiferromagnetic semiconductor to a Weyl semimetal hosting four symmetry-related Weyl points and robust Fermi arc states. Systematic substitution of Mn with Fe, Co, and Zn further reveals a tunable sequence of magnetic ground states: Fe and Co induce ferrimagnetism with semimetallic behavior, while Zn stabilizes a ferromagnetic semimetal with a large net moment. These findings establish Bi-based EuMnXBi$_2$ pnictides as a versatile platform where magnetic exchange interactions and band topology can be engineered through SOC and chemical substitution. The complex interplay of magnetic interactions and topological effects in the proposed bulk and doped pnictides opens a promising avenue to explore a wide range of electronic and magnetic phenomena. In particular, this study demonstrates that EuMn$_2$Bi$_2$ hosts tunable magnetic and topological phases driven by electron correlations, chemical substitution, and spin-orbit coupling.
\end{abstract}




\maketitle
\newpage
\section{\label{sec:level1}Introduction}

Layered pnictides with the general formula $AB_{2}P_{2}$ ($A$ = rare-earth or alkaline-earth metal, $T$ = transition metal, $P$ = pnictogen) have long served as a rich platform to study correlated electronic states, magnetism, and emergent topological phases. Compounds crystallizing in the tetragonal ThCr$_2$Si$_2$-type~\cite{tegel2008structural,stewart2011superconductivity} and trigonal CaAl$_2$Si$_2$-type~\cite{ruhl1979new} structures exhibit diverse quantum phenomena, including high-$T_c$ superconductivity, spin-density-wave transitions, and magnetically driven metal-insulator transitions. The ThCr$_2$Si$_2$-type iron pnictides, such as EuFe$_2$As$_2$, have been extensively explored for their pressure- and doping-induced superconductivity~\cite{zapf2016europium,miclea2009evidence}. The CaAl$_2$Si$_2$-type structure is mainly driven by its topological and thermoelectric properties. For example, both electrons and holes contribute to the electrical transport properties of the prototype material CaAl$_2$Si$_2$~\cite{imai2004electrical,krishna2018first}. Within the CaAl$_2$Si$_2$-type structure, the transition metal can have fully occupied ($d^{10}$), partially filled ($d^{5}$), or unoccupied ($d^{0}$) $d$ orbitals. In contrast, in the ThCr$_2$Si$_2$-type structure, there is no such preference for the transition metal~\cite{anand2016metallic}. 

Recently, researchers have turned their attention towards layered trigonal Mn-based pnictides. This shift arises due to the challenges associated with Fe-based pnictides, which are high-$T_c$ superconductors but face significant difficulties in achieving the desired transition temperatures ($T_c$'s)~\cite{simonson2012magnetic}. Most Mn-based layered pnictides exhibit the properties of antiferromagnetic insulators~\cite{simonson2012magnetic}. The induction of an electronic delocalization transition (EDT)~\cite{basov2011electrodynamics} through doping or applying pressure could potentially transform them into a strongly correlated metallic state, which is a crucial precursor for unconventional superconductivity. However, compounds such as LaMnPO~\cite{simonson2011gap} and BaMn$_2$As$_2$~\cite{pandey2011large} remain insulators even after heavy doping. Consequently, despite the EDT induced by doping or pressure, superconductivity has not been observed in these cases. 

The Mn-based $AB_{2}P_{2}$ pnictides typically stabilize in robust antiferromagnetic (AFM) insulating states~\cite{simonson2011gap}, and attempts to induce metallicity or superconductivity via chemical substitution or external pressure have largely been unsuccessful~\cite{simonson2011gap,dahal2019spin}. Among the EuMn$_2$X$_2$ ($X$ = P, As, and Sb) family, various AFM transitions and semiconducting behaviors have been reported. These pnictide compounds were first proposed in 1979 by R. Ruhl \textit{et al.}~\cite{ruhl1979new}, who introduced rare-earth and transition-metal Mn-based pnictides crystallizing in either anti-Ce$_2$O$_2$S or CaAl$_2$Si$_2$-type structures. Since then, many studies have investigated the structural and magnetic properties of the EuMn$_2$X$_2$ series. Anand \textit{et al.}~\cite{anand2016metallic} found that EuMn$_2$As$_2$ possesses an insulating ground state with an activation energy of around $53~meV$. Furthermore, it demonstrates antiferromagnetic (AFM) ordering of Eu$^{2+}$ moments with $S = 7/2$ at both $15$~K and $5$~K, while the Mn$^{2+}$ moments exhibit AFM ordering with $S = 5/2$ at $142~K$. When doped with 4\% and 7\% potassium (K) for Eu in Eu$_{1-x}$K$_x$Mn$_2$As$_2$, the material transitions to a metallic ground state. Dahal \textit{et al.}~\cite{dahal2019spin} reported that EuMn$_2$As$_2$ displays two distinct AFM transitions on $T = 135$~K and $14.4$~K. The AFM behavior of EuMn$_2$P$_2$ with a N\'eel temperature around $16.5$~K, was discovered by Payne \textit{et al.}~\cite{payne2002synthesis}. Between $300$~K and $120$~K, this compound exhibits semiconducting behavior with an activation energy of $38.9~meV$. Berry \textit{et al.}~\cite{berry2023bonding} discovered that EuMn$_2$P$_2$ remains an AFM insulator at all temperatures, undergoing a phase transition at $T = 20$~K for Eu long-range AFM ordering and at $T = 130$~K for Mn long-range AFM ordering. With an indirect band gap of roughly $450~meV$ for Mn magnetic ordering and Eu in the core, the compound also demonstrates semiconducting behavior. However, considering the Eu magnetic order, the indirect gap significantly decreases to around 0.29~eV. Schellenberg \textit{et al.}~\cite{schellenberg2010121sb} reported that the magnetic moment of EuMn$_2$Sb$_2$ is zero, and it displays AFM behavior with the magnetic moments of Eu$^{2+}$ and Mn$^{2+}$ being 7.94~$\mu_B$ and 5.92~$\mu_B$, respectively. While As-, Sb-, and P-based Eu--Mn pnictides are well studied, little attention has been paid to the Bi-based system. Karl \textit{et al.}~\cite{gschneidner2004handbook} reported that all EuMn$_2$X$_2$ (X = As, Sb, Bi, P) compounds may be isostructural and possibly semiconducting. Very recently, Khatri \textit{et al.}~\cite{khatri2023advancing} predicted magnetization in several Mn-based pnictide compounds using machine learning techniques and reported the stability of these, including Bi-based pnictides, based on the formation-energy criterion. However, their Bi-based analogue EuMn$_2$Bi$_2$ has received little attention, despite theoretical predictions based on machine learning techniques indicating its stability according to the formation-energy criterion~\cite{khatri2023advancing}. Recently, \cite{choudhury2024emerging} predicted from first-principle calculations that EuMn$_2$Bi$_2$ hoasts a C$_0$-AFM ground state with a very small band gap and several nearly degenerate magnetic states exhibiting Dirac/Weyl semimetal and topological insulating behavior. This establishes EuMn$_2$Bi$_2$ as a promising platform for magnetic topological phases. However, the roles of spin-orbit coupling (SOC), magnetic exchange interactions, and transition-metal substitution in stabilizing these phases remain unexplored.

This gap is particularly significant because bismuth contributes strong $6p$ spin-orbit coupling (SOC), which can drive band inversions and topological phase transitions when combined with broken time-reversal symmetry from magnetic order. Such conditions are known to stabilize magnetic Weyl semimetal (WSM) phases in related systems such as EuCd$_2$Bi$_2$ and Co$_3$Sn$_2$S$_2$~\cite{wang2021magnetic,grassano2024type}. Uncovering similar behavior in EuMn$_2$Bi$_2$ could provide a rare platform to tune magnetism and topology within a single material system. 

Here, we present a comprehensive density functional theory (DFT) study of EuMn$_2$Bi$_2$ and its doped analogues EuMnXBi$_2$ ($X$ = Fe, Co, Zn). Our calculations reveal that pristine EuMn$_2$Bi$_2$ stabilizes in a C-type AFM ground state and undergoes an SOC-driven transition to a WSM hosting four symmetry-related Weyl points and topological Fermi arc surface states. Substituting Mn with Fe or Co indues ferrimagnetism with semimetallic behavior, while Zn stabilizes a ferromagnetic semimetal with a large net moment. 

These results establish EuMnXBi$_2$ as a tunable platform where magnetic exchange interactions and band topology can be engineered via chemical substitution and SOC, providing new opportunities for realizing correlated magnetic topological phases relevant to next-generation spintronic devices.

\section{\label{sec:level2}Computational Methods}

First-principle calculations were performed using the \textsc{VASP} package~\cite{blochl1994projector,lopez2016first} within the framework of density functional theory (DFT) and the projector-augmented wave (PAW) method. The exchange-correlation energy was treated using the generalized gradient approximation (GGA) of Perdew, Burke, and Ernzerhof (PBE)~\cite{kresse1999ultrasoft}. A plane-wave energy cutoff of 520~eV and a Gaussian smearing width of 0.05~eV were used, ensuring total-energy convergence within 1~meV per formula unit. Structural relaxations employed a Monkhorst-Pack $5\times5\times3$ $k$-point mesh, while electronic density of states (DOS) calculations used a denser $10\times10\times6$ mesh~\cite{pack1977special}.

Strongly correlated Eu-$4f$ and Mn-$3d$ states were treated using the GGA+$U$ scheme~\cite{schlipf2013structural,chen2024electronic} with effective Hubbard parameters $U_{\mathrm{eff}}=U-J$ of 6.32~eV for Eu and 4.16~eV for Mn, determined from the linear response method~\cite{cococcioni2005linear}. Additional calculations were also carried out with $U_{\mathrm{eff}}=5$~eV for comparison~\cite{berry2023bonding}. Spin-orbit coupling (SOC) was included self-consistently, as it is expected to significantly influence the band topology of these Bi-based compounds~\cite{nguyen2020spin}.

Topological properties of \(\mathrm{EuMn_2Bi_2}\), including Berry curvature, Weyl points, and surface states, were evaluated using \textsc{Wannier90}~\cite{mostofi2014updated} to construct maximally localized Wannier functions (MLWFs)~\cite{marzari2012maximally}, and \textsc{WannierTools}~\cite{wu2018wanniertools} to compute Berry curvature distribution, surface spectral functions, and Fermi arcs from the resulting tight-binding Hamiltonian.

Magnetic exchange interactions were modeled using the Heisenberg Hamiltonian
\begin{equation}
H = -\sum_{i,j}J_{ij}\,\mathbf{S}_i \cdot \mathbf{S}_j,
\end{equation}
where $J_{ij}$ represents the effective exchange coupling between spins $i$ and $j$. exchange parameter was extracted from the total-energy difference $\Delta E = E_{\mathrm{AFM}} - E_{\mathrm{FM}}$ as
\begin{equation}
J_{\mathrm{eff}} = \frac{\Delta E}{2 N Z S^2},
\end{equation}
where $N$ is the number of magnetic atoms, $Z$ the number of nearest neighbors, and $S$ the spin quantum number.
The Curie temperature was then estimated using the molecular-field approximation:
\begin{equation}
T_c = \frac{2}{3} \, \frac{Z J S (S+1)}{k_B},
\end{equation}
where $k_B$ is the Boltzmann constant~\cite{coey2010magnetism,naji2014electronic,naji2014adsorption,el2013first}.

\begin{figure*}
    \centering
    \includegraphics[width=1\linewidth]{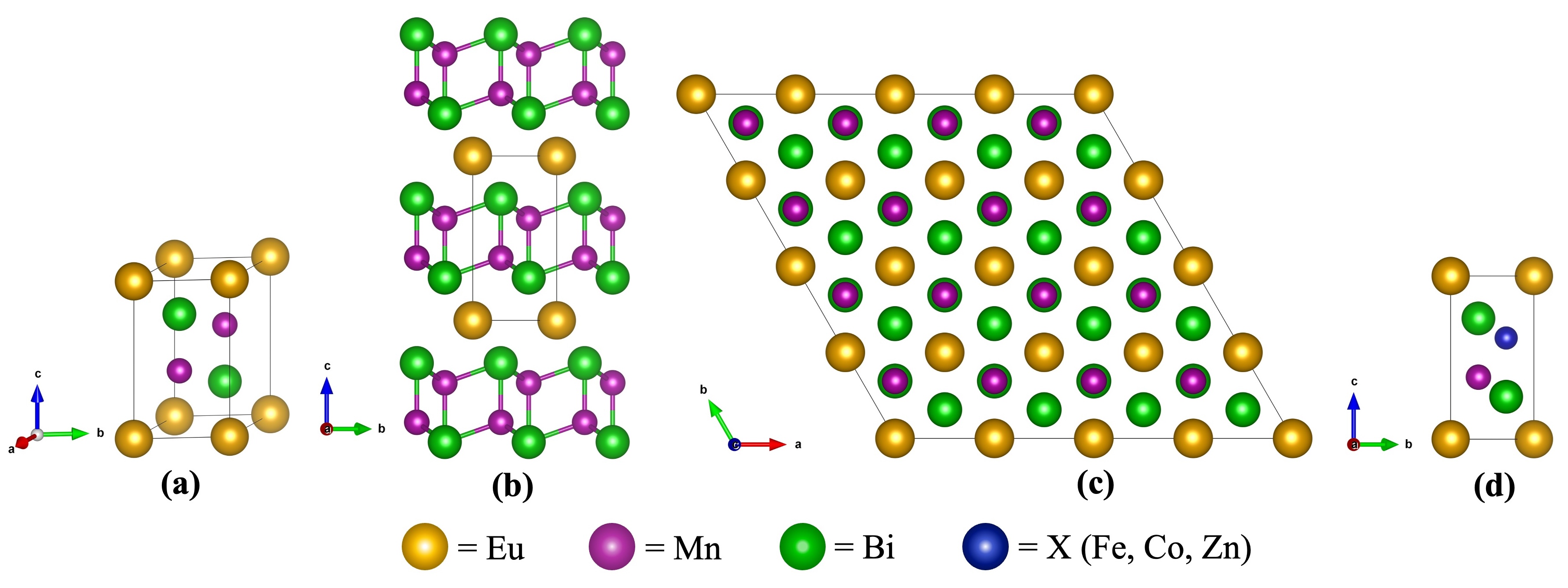}
    \caption{Crystal structure of EuMn$_2$Bi$_2$, EuMnXBi$_2$ (X = Fe, Co, and Zn), and arrangement of Eu, Mn, and Bi atoms. (a) Crystal structure of EuMn$_2$Bi$_2$, (b) the [MnBi$_2$]$^{-2}$ networks and Mn--Bi layers. (c) top view showing triangular Eu and honeycomb Mn sublattices. (d) Unit cell of Fe, Co, and Zn replacing X in EuMnXBi$_2$. yellow = Europium (Eu), purple = Manganese (Mn), and green = Bismuth (Bi), and blue = Iron (Fe), Cobalt (Co), and Zinc (Zn).}
    \label{fig:placeholder}

\end{figure*}
\begin{figure}
    \centering
    \includegraphics[width=1\linewidth]{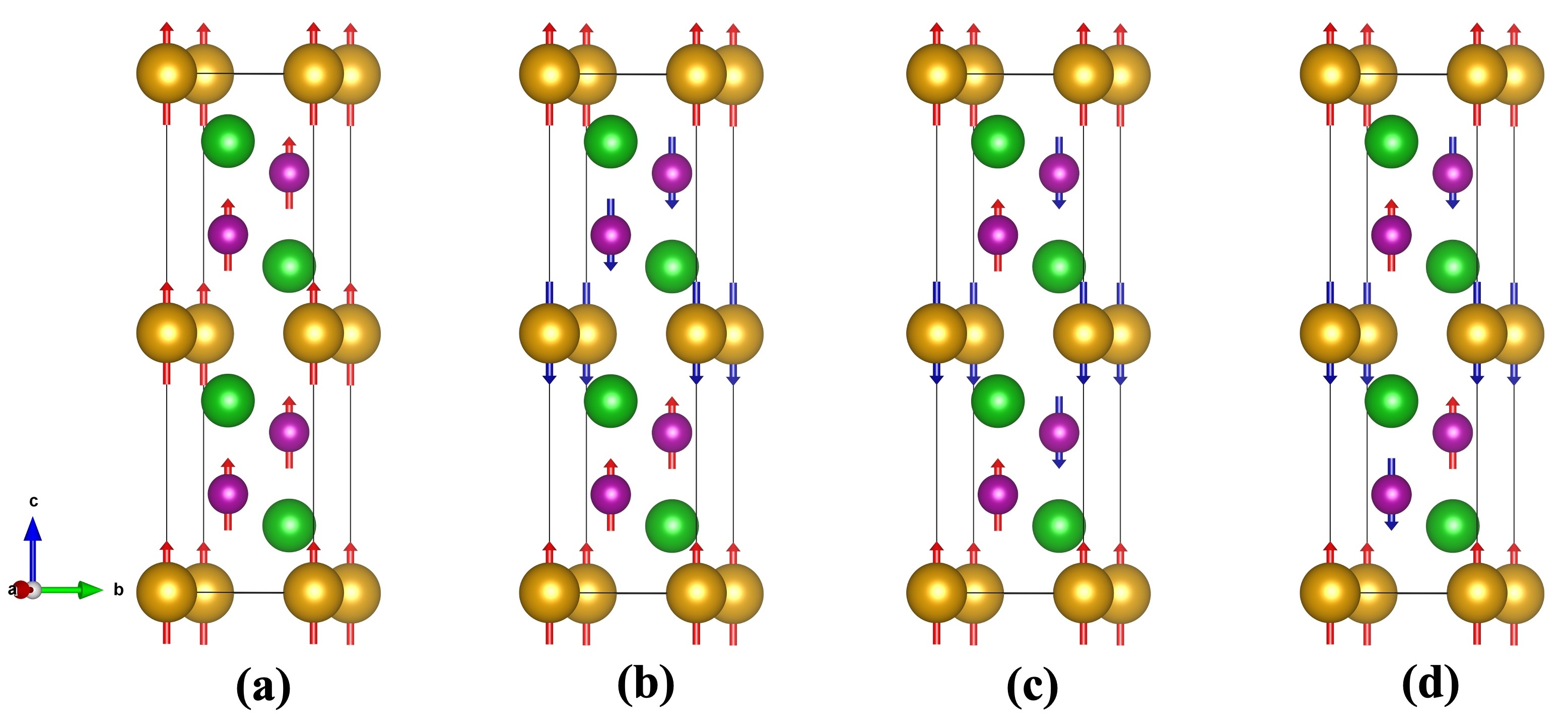}
    \caption{Magnetic crystal structure of the EuMn$_2$Bi$_2$ supercell ($1 \times 1 \times 2$) for Ferromagnetic and three types of anti-ferromagnetic magnetic configurations. (a) FM. (b) A-AFM. (c) C-AFM. (d) G -AFM.}
    \label{fig:placeholder}
\end{figure}

\FloatBarrier  
\begin{table}[H]
\centering
\setlength{\tabcolsep}{8pt} 
\renewcommand{\arraystretch}{1.0} 
\caption{Calculated lattice parameters of EuMn$_2$Bi$_2$ and doped compounds.}
\label{tab:lattice}
\begin{tabular}{lcccc}
\toprule
Compound & $a$ (\AA) & $c$ (\AA) & $c/a$ & $V$ (\AA$^{3}$) \\
\midrule
EuMn$_2$Bi$_2$  & 4.618 & 7.884 & 1.71 & 145.64 \\
EuMnFeBi$_2$    & 4.384 & 8.109 & 1.85 & 134.99 \\
EuMnCoBi$_2$    & 4.238 & 8.320 & 1.96 & 129.46 \\
EuMnZnBi$_2$    & 4.684 & 7.729 & 1.65 & 146.91 \\
\bottomrule
\end{tabular}
\end{table}

\newpage
\section{\label{sec:level2}Results and discussion }

\subsection{\label{sec:citeref}\textit{Crystal structure optimization and magnetic ordering}}
EuMn$_2$Bi$_2$ crystallizes in the trigonal CaAl$_2$Si$_2$-type structure (space group $P\bar{3}m1$, $No.~164$). Eu occupies the 1$a$ (0,0,0) site, while Mn and Bi occupy 2$d$ (1/3,2/3,0.62) and (1/3,2/3,0.62) sites, respectively [Fig.~1(a)]. Fig.~1(b) illustrates the EuMn$_2$Bi$_2$ structure, highlighting the [MnBi$_2$]$^{-2}$ networks and Mn--Bi layers, whereas Fig.~1(c) shows the hexagonal arrangement of Eu, Mn, and Bi atoms in the EuMn$_2$Bi$_2$ structure. The Eu atoms form triangular-lattice layers, while the Mn atoms form a corrugated honeycomb-like lattice. The unit cell for the substituted system is presented in Fig.~1(d). 

Relaxed lattice constants are $a = 4.618$~\AA\ and $c = 7.884$~\AA ($V=145.64$~\AA$^3$), consistent with other EuMn$_2$X$_2$ pnictides~\cite{anand2016metallic,payne2002synthesis,berry2023bonding,schellenberg2010121sb}. Doped EuMnXBi$_2$ ($X$=Fe, Co, Zn) compounds exhibits comparable equilibrium volume to pristine EuMn$_2$Bi$_2$, indicating their structural stability. The fully relaxed lattice parameters of EuMn$_2$Bi$_2$ and its doped analogues  (obtained by substituting one Mn atom in the EuMn$_2$Bi$_2$ unit cell) are summarized in Table~1.

For all the systems, EuMnXBi$_2$ (X = Mn, Fe, and Co), the Eu, Mn, and X spin moments were considered in both ferromagnetic (FM) and antiferromagnetic (AFM) arrangements. In the FM configuration, the spin orientations of the Eu, Mn, and X atoms are aligned $z-$axis, whereas in the AFM configuration, three distinct spin arrangements emerge: A-AFM, C-AFM, and G-AFM. In all AFM configurations, the Eu spin moments are orientated along the $z-$axis and cancel each other between the top and bottom planes. In the EuMn$_2$Bi$_2$ structure, the A-AFM configuration is characterized by Mn spin moments that cancel each other between the top and bottom planes, as shown in Fig.~2(b).  The C-AFM configuration is characterised by Mn spin moments that cancel each other over parallel planes, as shown in Fig.~2(c).  The G-AFM configuration is characterised by Mn spin moments that cancel each other both between the top and bottom planes and between parallel planes, as shown in Fig.~2(d).  In EuMnXBi$_2$ (X = Fe, Co), the spin configuration is consistent in both the FM state and the three AFM configurations, whereas in EuMnZnBi$_2$, the FM and A-AFM configurations are exits. Due to the extremely low magnetic moments of Bi and Zn, their spins were not considered in our calculations.

\subsection{\label{sec:citeref}\textit{Electronic and magnetic properties of antiferromagnetic EuMn$_2$Bi$_2$}}

The total energy calculations for both the ferromagnetic (FM) and various antiferromagnetic (AFM) configurations were performed within GGA and GGA+$U$ using different values of the Hubbard $U$, and the results are summarized in Table~2. To provide a better description of the partially filled and strongly correlated electronic states in the highly localized rare-earth Eu-4$f$ and transition-metal Mn-3$d$ orbitals, the GGA+$U$ method was adopted~\cite{schlipf2013structural,chen2024electronic}. We used two different values of the effective Hubbard parameter, $U_{\mathrm{eff}} = U - J$, for the half-filled Eu-$f$ orbital and the partially filled Mn-$d$ orbital, where $U$ is the on-site Coulomb correlation and $J$ is Hund’s exchange. First, we employed $U_{\mathrm{eff}} = 5$~eV for both Eu and Mn, as reported in Ref.~\cite{berry2023bonding}. Second, we used $U_{\mathrm{eff}}$ values of 6.32~eV and 4.16~eV for Eu and Mn, respectively, which we calculated \textit{ab-initio} using the Linear Response (LR) approach~\cite{cococcioni2005linear}. This approach has proven to be more reliable than the empirical approach~\cite{himmetoglu2014hubbard,mahajan2021importance}, thereby avoiding the empiricism and ambiguities common in many DFT+$U$ studies. Notably, the \textit{ab-initio} value of $U_{\mathrm{eff}} = 6.32$~eV obtained for Eu closely matches the $U_{\mathrm{eff}} = 6.3$~eV value employed in EuCd$_2$As$_2$~\cite{li2021engineering}.

In GGA, the lowest energy state of EuMn$_2$Bi$_2$ is found to be the G-AFM configuration, with the FM and A-AFM states lying much higher in energy compared to the C-AFM and G-AFM states. The lowest energy state within GGA+$U$ ($U_{\mathrm{eff}} = 5$~eV for both Eu and Mn) is also G-AFM; however, the C-AFM state is nearly degenerate in energy. When the values of $U_{\mathrm{eff}}$ are 6.32~eV and 4.16~eV for Eu and Mn, respectively, the lowest energies state of the C-AFM and G-AFM states becomes nearly identical, as shown in Table~2. This indicates that C-AFM and G-AFM are equi-energetic magnetic states, both lying lower in energy than the other magnetic configurations. Consequently, detailed calculations were carried out only for the C-AFM state. The extremely small energy separation between the C-AFM and G-AFM states indicates that EuMn$_2$Bi$_2$ resides very close to a magnetic phase boundary. Such magnetic near-degeneracy makes the system highly responsive to external perturbations, including chemical substitution, lattice strain, hydrostatic pressure, and applied magnetic fields. Even modest perturbations may therefore alter the balance between the two AFM configurations and potentially drive a transition between C-AFM and G-AFM order. Similar sensitivity to external tuning has been widely reported in correlated and frustrated magnets hosting nearly degenerate spin states~\cite{moessner2006geometrical,vzutic2004spintronics,foyevtsova2013ab,rau2016spin}. In EuMn$_2$Bi$_2$, this is particularly relevant because the G-AFM state has been predicted to host a Dirac semimetal phase~\cite{choudhury2024emerging}, while our LR-derived Hubbard $U$ values stabilize the C-AFM state, which lies $6~meV$ lower in energy than G-AFM. The effect of this near-degeneracy becomes even more evident upon doping. The substitution of Fe/Co/Zn perturbs the Mn–Bi–Mn super-exchange pathways and modifies the local electronic environment Mn-3$d$, thus lifting the intrinsic energy balance and robustly stabilizing the C-AFM order, further details are provided in Section 3.4. 

Using GGA+$U$, the energy difference between AFM and FM configurations is $-0.37~eV$. Our system gives $J_{\mathrm{eff}} = -1.64$~meV and $-1.1$~meV, indicating strong exchange interactions $J_{\mathrm{Mn-Mn}}$ and $J_{\mathrm{Eu-Mn}}$~\cite{khatri2021magnetic}, whereas the exchange interaction $J_{\mathrm{Eu-Eu}}$ is much weaker, with a value of $-0.08$~meV. The Curie temperature $T_c$ was estimated using the molecular-field approximation [Eq.(3)], yielding $T_c = 278.5$~K. This value is known to systematically overestimate ordering temperature because spin fluctuations are neglected~\cite{rusz2006exchange,tran2025realistic}.

Further, the magnetic properties of EuMn$_2$Bi$_2$ were investigated using both GGA and GGA+$U$ methods. Experimental and theoretical data for Eu$^{2+}$ ions indicate magnetic moments of $7.89~\mu_B$/Eu and $7~\mu_B$/Eu, respectively, with spin $S = 7/2$, exhibiting AFM ordering~\cite{anand2016metallic}. For Mn$^{2+}$ ions, the magnetic moment is reported to be $5.92~\mu_B$/Mn, also exhibiting AFM ordering~\cite{schellenberg2010121sb}. In our study, the calculated magnetic moments for Eu$^{2+}$ ions are $\pm 6.811~\mu_B$/Eu, $\pm 6.964~\mu_B$/Eu, and $\pm 6.982~\mu_B$/Eu in the GGA and GGA+$U$ methods, respectively. For Mn$^{2+}$ ions, the magnetic moments are $\pm 3.946~\mu_B$/Mn, $\pm 4.619~\mu_B$/Mn, and $\pm 4.554~\mu_B$/Mn in the GGA and GGA+$U$ methods, respectively, as shown in Table~2. The total magnetic moment of EuMn$_2$Bi$_2$ is found to be zero, consistent with AFM magnetic ordering.

\begin{table}[h]
\centering
\setlength{\tabcolsep}{10pt} 
\renewcommand{\arraystretch}{1.0} 
\caption{Magnetic configurations of EuMn$_2$Bi$_2$ calculated within GGA, 
GGA+$U$ ($U_\text{Eu}=5.00$ eV, $U_\text{Mn}=5.00$ eV) and 
GGA+$U$ ($U_\text{Eu}=6.32$ eV, $U_\text{Mn}=4.16$ eV). 
Listed are the total energy difference $\Delta E = E_{\mathrm{AFM}}-E_{\mathrm{FM}}$ (eV/f.u.), 
band gap $E_g$ (eV), and site-projected magnetic moments ($\mu_B$).}
\label{tab:mag}
\begin{tabular}{lccc}
\toprule
 & GGA & \multicolumn{2}{c}{GGA+U} \\
\midrule
$U_{\mathrm{eff}}$ (Eu)   & ---    & 5.00  & 6.32 \\
$U_{\mathrm{eff}}$ (Mn)   & ---    & 5.00  & 4.16 \\
\midrule
FM       & $-72.050$ & $-65.271$ & $-65.920$ \\
A-AFM    & $-72.060$ & $-65.333$ & $-65.981$ \\
C-AFM    & $-73.030$ & $-65.588$ & $-66.296$ \\
G-AFM    & $-73.043$ & $-65.596$ & $-66.290$ \\
$\Delta E$ & $-0.993$  & $-0.320$ & $-0.370$ \\
\midrule
$E_g$        & 0.0     & 0.122  & 0.258 \\
$\mu_{\mathrm{Eu}}$ & 6.81  & 6.96  & 6.98 \\
$\mu_{\mathrm{Mn}}$ & 3.94  & 4.62  & 4.55 \\
\bottomrule
\end{tabular}
\end{table}

As heavy orbitals such as Eu-4$f$, Mn-3$d$, and Bi-6$p$ are present in this compound, the effect of spin-orbit coupling (SOC) is significant, and SOC plays an important role in calculating the electronic properties of the material~\cite{nguyen2020spin}. Therefore, SOC was included in the electronic structure calculations. We observe that the GGA+$U$+SOC results are superior when compared to those obtained from GGA and GGA+$U$, since the ground-state energy obtained with SOC is lower than that with GGA+$U$, and the exchange energy is also reduced compared to GGA and GGA+$U$, as shown in Table~2.

\begin{figure}
    \centering
    \includegraphics[width=1\linewidth]{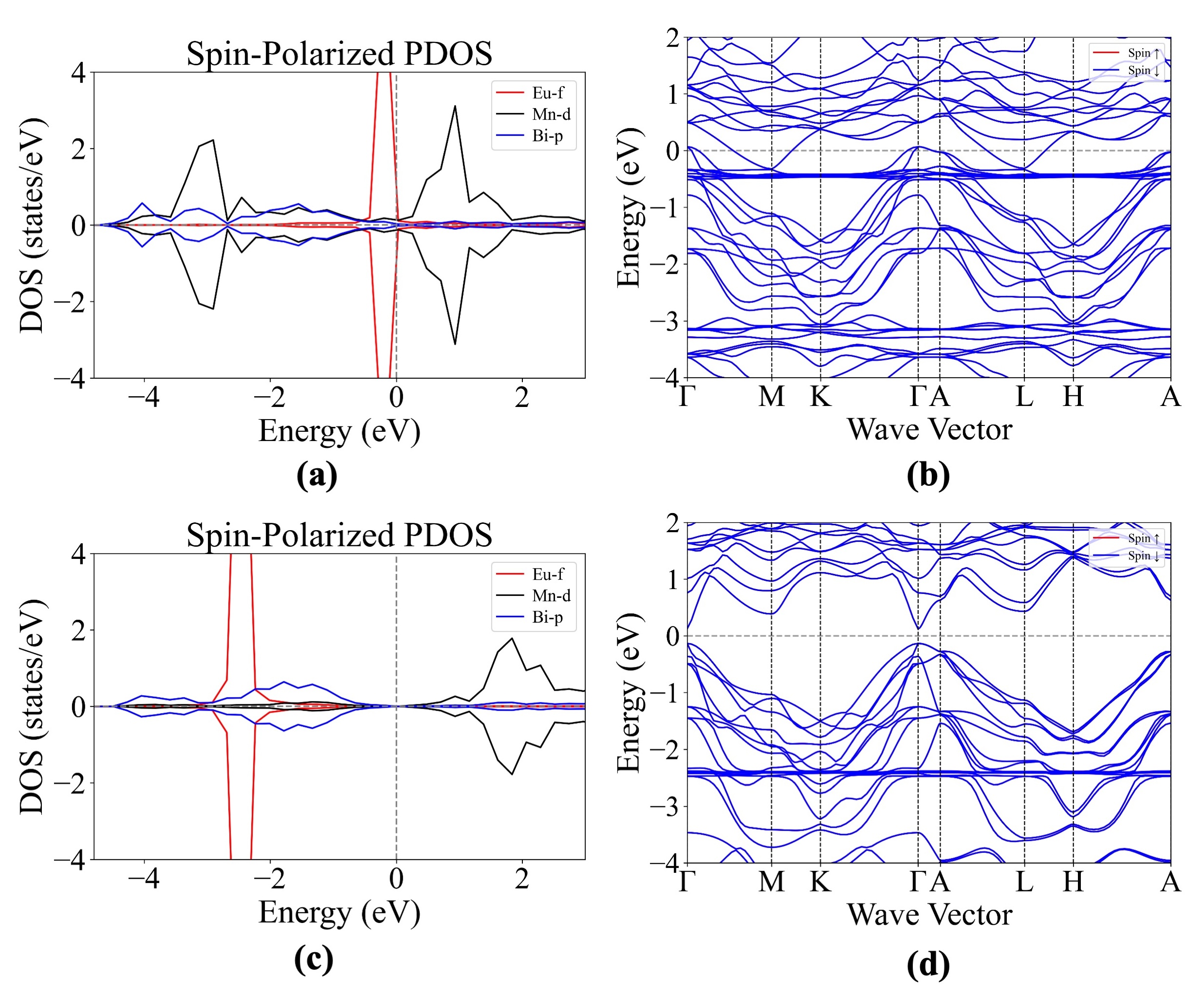}
    \caption{Spin-polarized projected density of state (PDOS) and electronic band structure for the magnetic ground state of EuMn$_2$Bi$_2$ in the G-AFM magnetic ground state (a)-(b) calculated within GGA. (c)-(d) calculated within GGA+U with $U_{eff} = 6.32~eV$ for Eu and $4.16~eV$ for Mn.}
    \label{fig:placeholder}
\end{figure}

Fig.~3(a) displays the Spin-polarized projected density of state (PDOS) for the G-AFM configuration within GGA. In this PDOS plot, the Eu-$f$ ($4f^7$) and Mn-$d$ ($3d^5$) states are highly localized at the Fermi level ($E_F$), with small localized Bi-$p$ states intersecting $E_F$. Moreover, the Eu-$d$, Mn-$s$, and Bi-$s$ states are highly delocalized below $E_F$. From the PDOS plots in GGA, it is evident that the G-AFM magnetic configuration exhibits metallic behavior. However, the GGA+$U$ method with $U_{\mathrm{eff}} = 5$~eV and $6.32$~eV for the Eu-$f$ state, and $U_{\mathrm{eff}} = 5$~eV and $4.16$~eV for the Mn-$d$ state across FM and all AFM magnetic configurations, opens up a band gap in the DOS. The ground-state energy calculated using $U$ obtained from the LR approach~\cite{cococcioni2005linear} is lower than that calculated using $U$ values taken from the literature, and the resulting band gap is also larger, as shown in Table~2. Therefore, we focus our discussion on the results obtained with $U = 6.32$~eV and $4.5$~eV. The PDOS for the electronic ground state within GGA+$U$ is presented in Fig.~3(c).

Fig.~3(c) shows the PDOS for C-AFM configuration calculated using GGA+$U$. The Eu-4$f$, Mn-3$d$, and Bi-$p$ states are shifted farther from the Fermi level ($E_F$), exhibiting increased localization compared to GGA, and the band gap expands to approximately $0.258$~eV within GGA+$U$. The similarity of the up- and down-spin components in both the valence bands (VBs) and conduction bands (CBs) for the GGA+$U$ electronic ground state suggests that the compound exhibits antiferromagnetic ordering. 

Next, we calculated the band structure using both GGA and GGA+$U$ methods. Fig.~3(b) illustrates the band structure for the electronic ground state (G-AFM) within GGA, plotted along high-symmetry directions. In this Fig., a few bands intersect the Fermi level ($E_F$) without spin crossings, indicating a semimetallic nature with no discernible band gap. Additionally, the $f$ orbitals appear localized near $E_F$, approximately at 0.5~eV. Fig.~3(d) shows the band structure for the C-AFM electronic ground state within GGA+$U$. It is evident that no bands intersect $E_F$, and the band gap increases with larger $U$ values for Eu. Furthermore, the $f$ states are shifted approximately 1.75--2.5~eV below $E_F$. Within GGA+$U$, the calculated band gaps are $0.122$~eV and $0.258$~eV for different $U$ values, as listed in Table~2, indicating a semiconducting nature with a direct band gap along the $\Gamma$ direction. Such AFM semiconducting behavior has also been reported for related layered pnictides such as EuMnBi$_2$ and EuCd$_2$As$_2$ \cite{masuda2016quantum,rahn2018coupling}. Fig.~4(b) presents the band structure for the C-AFM electronic ground state within GGA+$U$+SOC. Here, the band touching at $E_F$ along the $\Gamma-A$ direction. Thus, 4.4 meV band gap is observed when SOC is included.

\begin{figure*}
    \centering
    \includegraphics[width=1\linewidth]{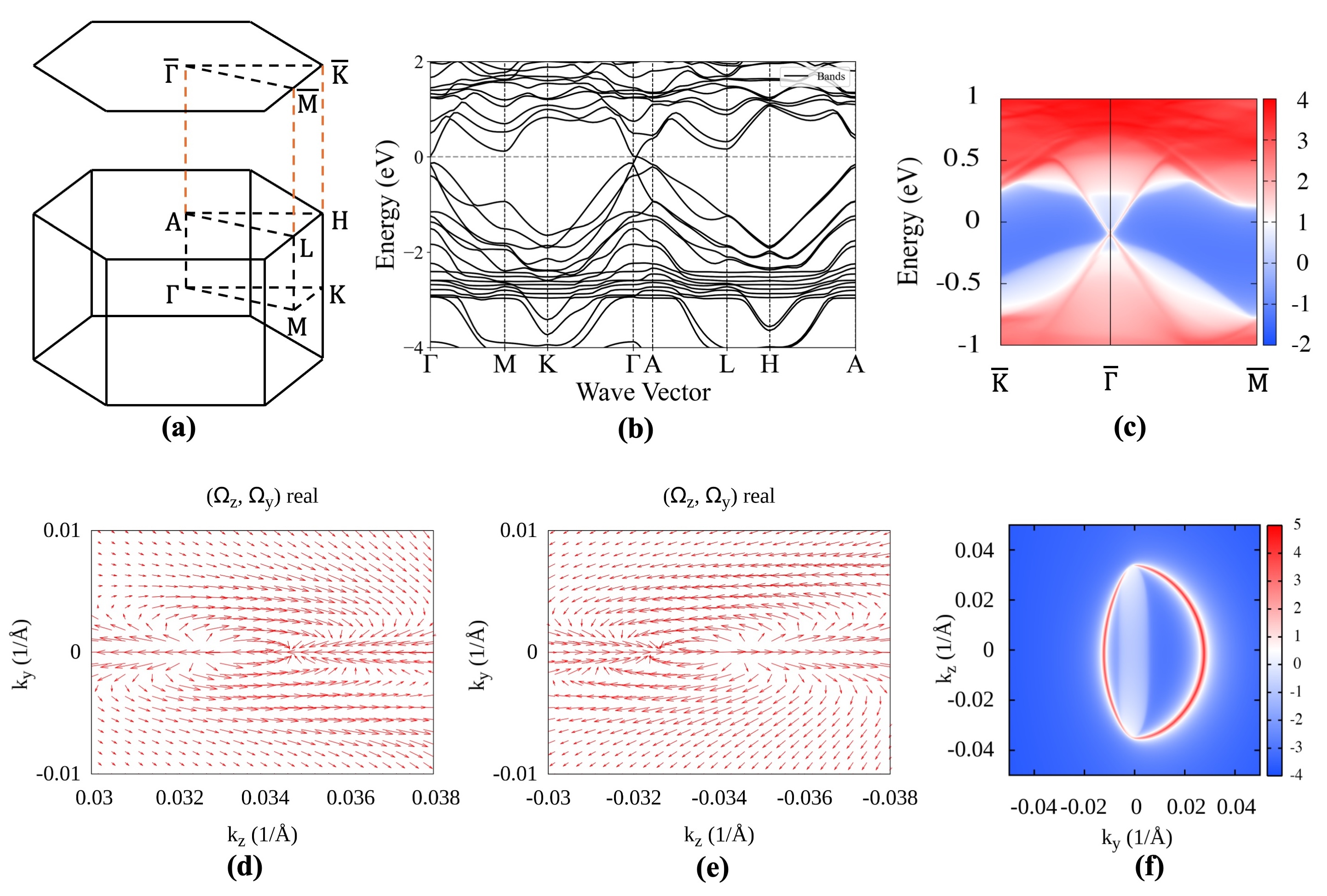}
    \caption{(a) Bulk Brillouin zone (BZ) and the corresponding (001) surface-projected BZ with high-symmetry points indicated. 
(b) Bulk electronic band structure of \(\mathrm{EuMn_2Bi_2}\) within SOC. (c) Surface spectral function of the (001) slab calculated along the \(\bar{K}\)-\(\bar{\Gamma}\)-\(\bar{M}\) direction. Berry curvature distribution around (d) WPs 1 and 2, and (e) WPs 3 and 4 (as listed in Table~3), showing the characteristic monopole-like source and sink patterns associated with opposite chiralities. (f) The surface-projected Fermi arcs that connect WPs of opposite chirality.}
    \label{fig:placeholder}
\end{figure*}

\begin{table}[h!]
\centering
\setlength{\tabcolsep}{4pt} 
\renewcommand{\arraystretch}{1.3} 
\caption{The primitive coordinates \((k_x, k_y, k_z)\) of two pairs of Weyl points (WPs) in the Brillouin zone (BZ) for \(\mathrm{EuMn_2Bi_2}\) in the C-AFM configuration are presented along with their corresponding energy \((E)\) and chirality \((C)\). The relative energies \((E)\) of the WPs are given with respect to the Fermi level.}
\label{tab:EuMn2Bi2}
\begin{tabular}{lcccccc}
\toprule
S. No. & $k_x$ & $k_y$ & $k_z$ & $E_{\mathrm{gap}}$ & $E$ & Chirality \\
       & (\AA$^{-1}$) & (\AA$^{-1}$) & (\AA$^{-1}$) & (eV) & (eV) & $(C)$ \\
\midrule
1 & 0 & 0 &  0.03366 & 0 & 0.018 &  1 \\
2 & 0 & 0 &  0.03341 & 0 & 0.018 & -1 \\
3 & 0 & 0 & -0.03341 & 0 & 0.018 &  1 \\
4 & 0 & 0 & -0.03368 & 0 & 0.018 & -1 \\
\bottomrule
\end{tabular}
\end{table}

\subsection{\label{sec:citeref}\textit{Topological properties}}

The total energy calculation place C-AFM and G-AFM states nearly degenerate, with an energy difference of only $6~meV/f.u$. Choudhury \textit{et al.}~\cite{choudhury2024emerging} have reported the G-AFM configuration to host a Dirac semimetal state. Therefore, in this section, we focus on the topological properties of the C-AFM configuration , which is marginally lower in energy and thus represent the most stable ground state. Topological states can often be identified by the presence of an inverted band structure arising from strong spin–orbit coupling (SOC) or crystal field effects \cite{armitage2018weyl,grassano2024type}. To examine this, we analyzed the bulk band structure and band symmetries of \(\mathrm{EuMn_2Bi_2}\) with and without SOC. As shown in Fig.~3(d), the C-AFM ground state exhibits an direct band gap of 0.258 eV near the $E_F$, indicating that the system behaves as an antiferromagnetic semiconductor.

\begin{figure}
    \centering
    \includegraphics[width=1\linewidth]{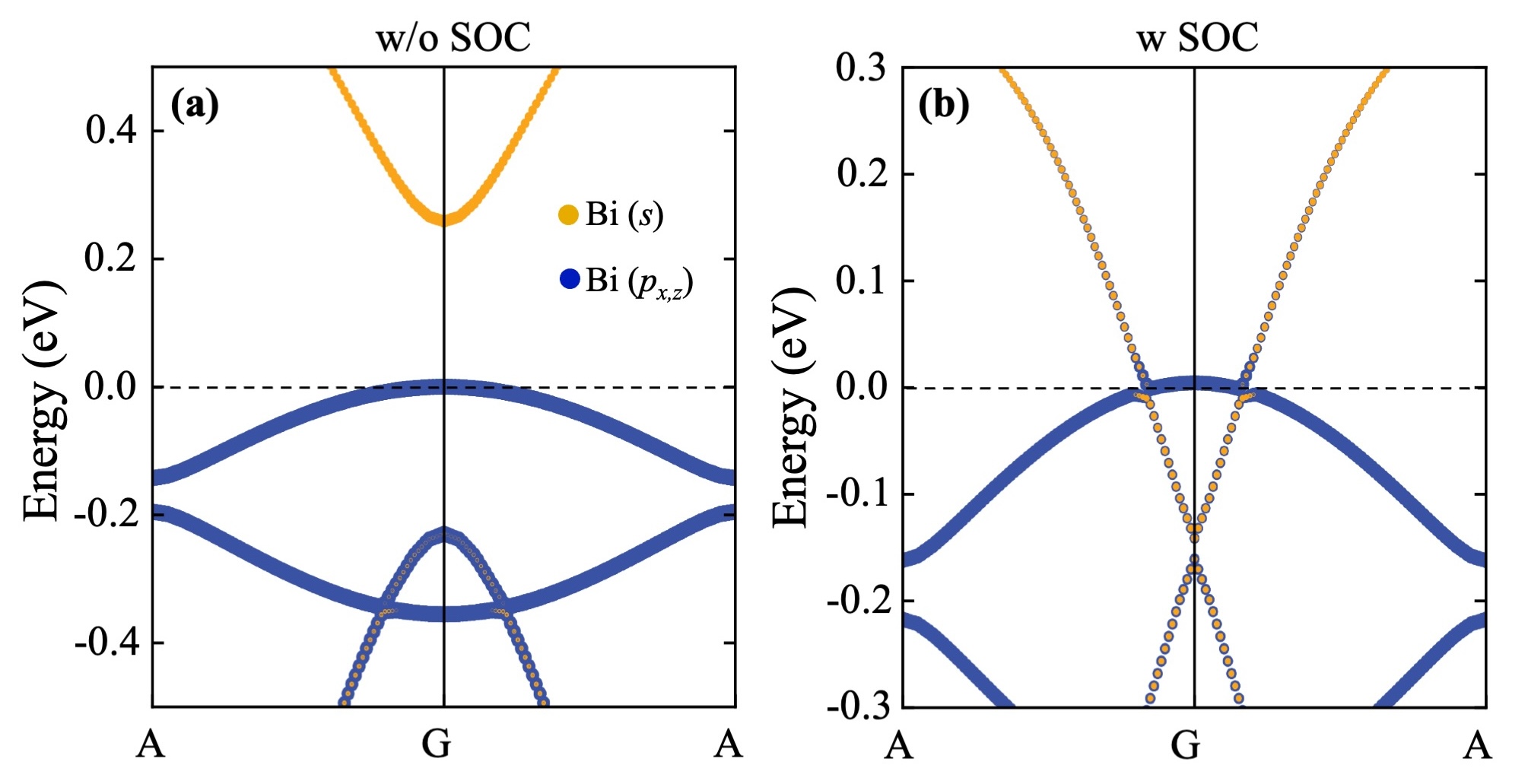}
    \caption{The band structures calculated (a) without and (b) with SOC, respectively. The projected contributions from the Bi $s$ and $p_{x,z}$ orbitals are also displayed.}
    \label{fig:placeholder}
\end{figure}
Because heavy orbitals such as Eu ($4f$), Mn ($3d$) and Bi ($6p$) are present, the effect of SOC is expected to be significant. When SOC is included in the C-AFM state, the band gap decrease to 4.4 meV and CBs band and VBs touch the \(E_F\) along the $\Gamma-A$ high-symmetry direction. This marks a topological phase transition from a trivial AFM semiconductor to a \textit{Weyl semimetal (WSM)}. A closer examination of the calculated band structures along the A–G–A high-symmetry path shows the evolution of the electronic states near the $E_F$ without and with SOC. In the absence of SOC [Fig.~5(a)], the Bi $p_{x,z}$ valence bands overlap near the E$_F$, while the Bi $s$ conduction band lies slightly above it. Upon inclusion of SOC, the relative ordering of these states is significantly modified, as the Bi $s$ band and the Bi $p_{x,z}$ bands cross the E$_F$, undergo a pronounced band inversion, and consequently reduce the band gap, as shown in Fig.~5(b). This SOC-driven reordering of orbital characters is consistent with the SOC-induced orbital inversion mechanisms discussed by Zhang \textit{et al.}~\cite{zhang2021selective}. A detailed search of the band structure found four symmetry-related Weyl points (WPs) located at ($0,0,0.03366$) and ($0,0,0.03341$), with corresponding symmetry-related-points at ($0,0,-0.03368$) and ($0,0,0.03341$), each WPs has a topological charge such as chirality \(C=\pm1\), present in Table~3. In this compound the coexistence of inversion symmetry (\(\mathcal{P}\)) and broken time-reversal symmetry (\(\mathcal{T}\)) enforces the Weyl points to appear in \(\pm\mathbf{k}\) pairs with opposite chirality, resulting in a net zero Chern number as required by the fermion doubling theorem \cite{armitage2018weyl,grassano2024type}.

To elucidate their topological nature, we calculated the Berry curvature \(\boldsymbol{\Omega}(\mathbf{k})\) throughout the BZ. The Berry curvature field around the WPs exhibits monopole-like behaviour, acting as a \textit{source} at the \(+1\) chirality nodes and a \textit{sink} at the \(-1\) chirality nodes as shown in Figs.~4(d)-(e). Such Berry curvature hotspots are the origin of many exotic transport responses in WSMs, including the intrinsic anomalous Hall effect and chiral anomaly \cite{armitage2018weyl,wang2021magnetic}. The topological character is further explored by the surface electronic structure. As shown in Fig.~4(c), the surface spectral function of a (001) slab along the \(\bar{K}\)-\(\bar{\Gamma}\)-\(\bar{M}\) path shows gapless surface states terminating at the projected bulk Weyl points. Moreover, a constant-energy contour at \(E=E_F\) reveals open two Fermi arcs connecting each pair of the surface projections of WPs with opposite chirality [see Fig.~4(f)]. These Fermi arcs are a hallmark of EuMn$_2$Bi$_2$ compound has a Weyl semimetals and similar feartures have been reported in other magnetic WSMs such as \(\mathrm{EuCd_2Bi_2}\), and \(\mathrm{Co_3Sn_2S_2}\) \cite{wang2021magnetic,grassano2024type}. Moreover, the topological characteristics of EuMn$_2$Bi$_2$ can be placed in the broader context of magnetism-driven topological phases that have been extensively reported in layered magnetic materials. In particular, recent studies on MnBi$_2$Te$_4$ and related compounds have demonstrated that intrinsic antiferromagnetic order, layer-dependent magnetism, and strong spin--orbit coupling can generate axion insulator states, quantum anomalous Hall phases, and tunable Weyl band topology~\cite{liu2020robust, deng2020quantum, li2019intrinsic, zhang2019topological, zhang2021tunable, zhang2020abundant}. These investigations collectively underscore how band inversion, surface-state gap formation, and Berry-curvature engineering are highly sensitive to magnetic configuration and chemical composition, thereby supporting the observed topological tunability in EuMn$_2$Bi$_2$.

\subsection{\label{sec:citeref}\textit{Electronic and magnetic properties of EuMnXBi$_2$}}

The electronic and magnetic properties of EuMn$X$Bi$_2$ (\(X =\) Fe, Co, and Zn) were investigated using the GGA+$U$ and GGA+$U$+SOC approaches. Owing to the strong electronic correlations associated with the Eu-$f$, Mn-$d$, Fe-$d$, and Co-$d$ orbitals, the GGA+$U$ scheme was adopted, with the effective Hubbard parameters (\(U_{\mathrm{eff}}\)) determined using the LR approch~\cite{cococcioni2005linear} (Table~4). The parent compound EuMn$_2$Bi$_2$ crystallizes in the trigonal CaAl$_2$Si$_2$-type structure (space group \(P\bar{3}m1\)). To examine substitutional effects, one Mn atom was replaced by Fe, Co, or Zn, and the corresponding ground-state energies were computed (Table~4).

For EuMnFeBi$_2$, the lowest-energy configuration is a C-AFM state, lying substantially below the FM, A-AFM, and G-AFM configurations, with \(\Delta E = E_{\mathrm{AFM}} - E_{\mathrm{FM}} = -0.31\)~eV. The density of states (DOS) reveals highly localized Fe-$d$ and Bi-$p$ states near the Fermi level (\(E_F\)), while the Eu-$f$, Mn-$d$, and Bi-$s$ states are more delocalized and cross \(E_F\) [Fig.~6(a)]. The unequal spin distribution indicates a ferrimagnetic (FiM) ground state, with calculated magnetic moments of \(\pm 7.024~\mu_B\)/Eu, \(\pm 4.575~\mu_B\)/Mn, and \(\pm 3.326~\mu_B\)/Fe, yielding a net moment of \(1.016~\mu_B\)/f.u. The band structure within GGA+$U$ [Fig.~6(b)] shows multiple conduction bands (CBs) crossing \(E_F\) at the \(\Gamma\), K–M, A, and L $k$-points, with no valence bands (VBs) intersecting \(E_F\), confirming a semimetallic nature. In the C-AFM state with SOC [Fig.~6(c)], several CBs intersect \(E_F\) and some VBs touch or cross \(E_F\), eliminating the gap. The orbital states broaden above and below \(E_F\), confirming that EuMnFeBi$_2$ adopts a ferrimagnetic semimetallic ground state.

\begin{table}[h!]
\centering
\setlength{\tabcolsep}{6pt} 
\caption{Calculated Hubbard parameters $U_{\mathrm{eff}}$ (eV), total energies (eV) of EuMnXBi$_2$ for FM and three AFM configurations within GGA+U, 
energy differences $\Delta E = E_{\mathrm{AFM}}-E_{\mathrm{FM}}$ (eV), and magnetic moments ($\mu_B$/atom).}
\label{tab:EuMnXBi2}
\begin{tabular}{lccc}
\toprule
 & EuMnFeBi$_2$ & EuMnCoBi$_2$ & EuMnZnBi$_2$ \\
\midrule
$U_{\mathrm{eff}}$ (Eu)  & 7.20 & 5.25 & 6.20 \\
$U_{\mathrm{eff}}$ (Mn)  & 5.00 & 3.37 & 4.68 \\
$U_{\mathrm{eff}}$ (Fe, Co) & 4.00 & 4.82 & --- \\
\midrule
FM      & $-61.14$ & $-59.16$ & $-54.11$ \\
A-AFM   & $-61.16$ & $-59.11$ & $-54.08$ \\
C-AFM   & $-61.45$ & $-59.46$ & --- \\
G-AFM   & $-61.32$ & $-59.39$ & --- \\
$\Delta E$ & $-0.31$ & $-0.30$ & $0.02$ \\
\midrule
$\mu_{\text{Eu}}$  & 7.024 & 6.971 & 7.00 \\
$\mu_{\text{Mn}}$  & 4.575 & 4.437 & 4.595 \\
$\mu_{\text{Fe}}$  & 3.326 & ---   & --- \\
$\mu_{\text{Co}}$  & ---   & 2.160 & --- \\
\bottomrule
\end{tabular}
\end{table}

Similarly, EuMnCoBi$_2$ stabilizes in a C-AFM configuration with \(\Delta E = E_{\mathrm{AFM}} - E_{\mathrm{FM}} = -0.30\)~eV. The DOS indicates pronounced Co-$d$, and Bi-$p$ localization near \(E_F\), while the Eu-$f$, Mn-$d$, and Bi-$s$ states are delocalized [Fig.~6(e)]. The unequal spin polarization indicates FiM ordering, with magnetic moments of \(\pm 6.971~\mu_B\)/Eu, \(\pm 4.437~\mu_B\)/Mn, and \(\pm 2.160~\mu_B\)/Co, giving a total moment of \(2.025~\mu_B\)/f.u., larger than that of EuMnFeBi$_2$. The band structure for the C-AFM state within GGA+$U$ [Fig.~6(f)] shows four CBs intersecting \(E_F\) at \(\Gamma\), K–M, A, and L $k$-points without spin crossings, while no VBs intersect \(E_F\), confirming its semimetallic nature. The Eu-$f$ orbitals are more delocalized, lying between 2.5 and 3~eV, and exhibit spin asymmetry between the spin-up and spin-down channels. In the presence of SOC [Fig.~6(g)], several CBs and some VBs cross or touch \(E_F\), removing the band gap. The broadened orbital states above and below \(E_F\) confirm that EuMnCoBi$_2$ also exhibits a ferrimagnetic semimetallic character.

In contrast, EuMnZnBi$_2$ favors a FM ground state within GGA+$U$, as the energy difference \(\Delta E = E_{\mathrm{AFM}} - E_{\mathrm{FM}} = 0.02\)~eV is negligible. The DOS shows weak localization of the Eu-$f$, Mn-$d$, Zn-$s$, and Bi-$p$ states near \(E_F\), and the unequal spin polarization confirms FM ordering [Fig.~6(i)]. The calculated magnetic moments are \(\pm 7.00~\mu_B\)/Eu and \(\pm 4.595~\mu_B\)/Mn, producing a large net moment of \(11.497~\mu_B\)/f.u., consistent with ferromagnetism. The band structure [Fig.~6(j)] shows multiple CBs and VBs crossing \(E_F\) at the \(\Gamma\) and A $k$-points without spin crossings, indicating semimetallic behavior. With SOC [Fig.~6(k)], both CBs and VBs broaden and touch \(E_F\), closing the gap and confirming its ferromagnetic semimetallic character.

\begin{figure*}
    \centering
    \includegraphics[width=1\linewidth]{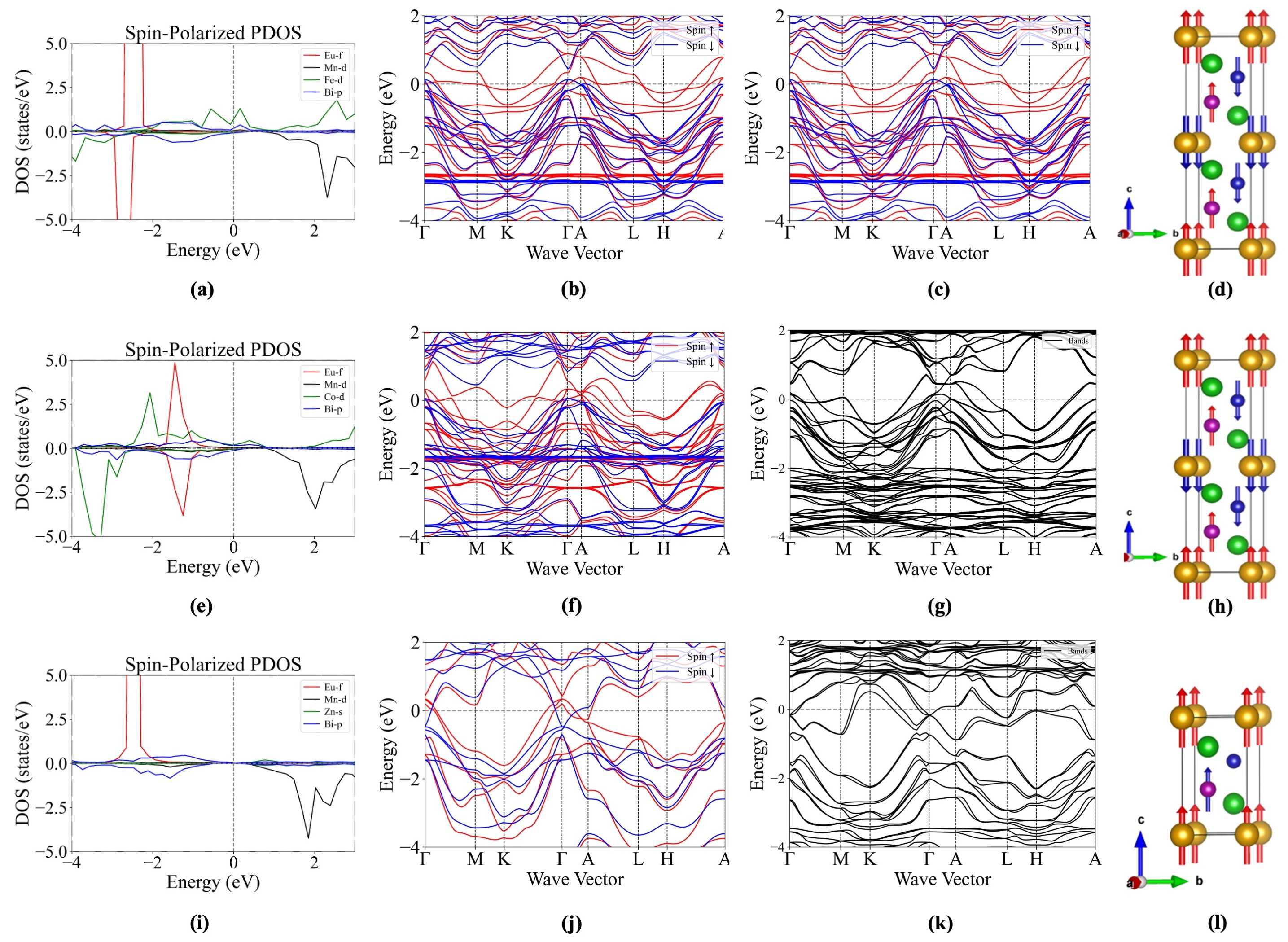}
    \caption{Spin-polarized projected density of state (PDOS), band structure, and crystal structure for magnetic ground state of the EuMnXBi$_2$ (X = Fe, Co, and Zn). (a)-(d) PDOS, band structure, and magnetic crystal structure of EuMnFeBi$_2$. (a)-(b) PDOS, along with the band structure for the C-AFM ground state within the GGA+U framework. (c) band structure for the C-AFM ground state incorporating GGA+U+SOC. (d) C-AFM crystal structure, with blue representing Fe atoms. Similarly, (e)-(h) PDOS, band structure, and magnetic crystal structure of EuMnCoBi$_2$. (e)-(f) PDOS, along with the band structure for the C-AFM ground state within the GGA+U framework. (g) band structure for the C-AFM ground state incorporating GGA+U+SOC. (h) C-AFM crystal structure, with blue representing Co atoms. Finally, (i)-(l) PDOS, band structure, and magnetic crystal structure of EuMnZnBi$_2$. (i)-(j) PDOS, along with the band structure for the FM ground state within the GGA+U framework. (k) band structure for the FM ground state incorporating GGA+U+SOC. (l) FM crystal structure, with blue representing Zn atoms. }
    \label{fig:placeholder}
\end{figure*}

To understand the origin of the distinct magnetic ground states in the substituted compounds, we analyzed the spin-polarized PDOS and orbital-resolved band structures of EuMnXBi$_2$ (X = Fe, Co, Zn), as shown in Fig.~6, together with the total-energy data in Table~4. The parent compound EuMn$_2$Bi$_2$ lies extremely close to a magnetic phase boundary, because the energy difference between C-AFM and G-AFM configurations is only a $6$~meV, indicating strong magnetic near-degeneracy. As a result, chemical substitution readily breaks this delicate balance by altering the Mn–Bi–Mn super-exchange pathways and the local Mn-3$d$ electronic environment, making doping an efficient tuning parameter. For EuMnFeBi$_2$ and EuMnCoBi$_2$, the Fe/Co 3$d$ states are partially occupied and carry sizable local moments of about 3.33 and 2.16~$\mu_B$, respectively, which couple antiparallel to the Mn moments ($\sim 4.5~\mu_B$), yielding a ferrimagnetic alignment within the Mn–X sublattice. The PDOS in Figs.~6(a) and 6(e) reveals strong hybridization between Fe/Co~3$d$ and Bi~$p$ states near the $E_F$, favoring antiferromagnetic Mn–X superexchange in accordance with the Goodenough–Kanamori rules~\cite{goodenough1958interpretation,kanamori1959superexchange}. Consistently, the C–AFM configuration is about 0.30–0.31~eV per formula unit lower in energy than the FM state for X = Fe and Co (Table~4), demonstrating that the parent compound’s near-degenerate AFM manifold is decisively tipped toward a ferrimagnetic ground state upon Fe or Co substitution.

In contrast, in EuMnZnBi$_2$ the Zn 4$s$ orbitals contribute only weakly near the $E_F$ [Fig.~6(i)], and Zn behaves as an almost nonmagnetic spacer. Here the doping effect is predominantly electronic: Zn substitution introduces additional electrons into the Mn–Bi network and shifts the Bi $p$ states closer to the \(E_F\) [Figs.~6(j) and 6(k)]. This enhances the itinerant Bi-$p$-mediated Mn–Mn ferromagnetic exchange, analogous to the electron-induced interlayer ferromagnetism reported in MnBi$_2$Te$_4$~\cite{zhang2025electron}. Consequently, EuMnZnBi$_2$ stabilizes a ferromagnetic ground state with a small positive energy difference, $\Delta E = E_{\mathrm{AFM}} - E_{\mathrm{FM}} \approx 0.02$~eV, highlighting how the parent compound’s magnetic near-degeneracy amplifies the system’s sensitivity to electron doping and leads to qualitatively different magnetic behavior for Fe/Co versus Zn substitution.

In summary, substitution of Mn by Fe or Co stabilizes ferrimagnetic ground states with semimetallic behavior, while Zn substitution drives the system into a ferromagnetic semimetallic state with a large net moment. These results demonstrate that transition-metal substitution in EuMn$_2$Bi$_2$ tunes the balance between AFM and FM exchange interactions, producing distinct magnetic ground states while retaining semimetallic character.

\section{\label{sec:level2}Conclusion}

The structural, electronic, magnetic, and topological properties of \(\mathrm{EuMn_2Bi_2}\) using density functional theory (DFT) within the GGA+\(U\) framework \((U_{\mathrm{Eu}} = 6.32~\mathrm{eV},\; \\U_{\mathrm{Mn}} = 4.16~\mathrm{eV})\). \(\mathrm{EuMn_2Bi_2}\) crystallizes in the trigonal CaAl\(_2\)Si\(_2\)-type structure with space group \(P\bar{3}m1\) (No.~164) and stabilizes in an antiferromagnetic ground state. The calculated magnetic moments are \(6.98~\mu_{\mathrm{B}}\)/Eu for Eu\(^{2+}\) and \(4.55~\mu_{\mathrm{B}}\)/Mn for Mn\(^{2+}\), consistent with previous experimental and theoretical results. The net magnetization is zero, consistent with AFM ordering. EuMn$_2$Bi$_2$ is classified as a narrow-gap semiconductor, with a direct band gap of 0.258~eV at the $\Gamma$ point. The Eu-\(f\) orbitals remain highly localized and lie well below the Fermi level (\(E_F\)), while the spin-up and spin-down bands are nearly symmetric, reflecting the compensated AFM state. Elemental substitution of Mn by Fe, Co, and Zn was also explored: Fe and Co doping drives the system toward a ferrimagnetic (FiM) state, while Zn doping stabilizes a ferromagnetic (FM) state, and these substituted systems exhibit semimetallic characteristics. The Mn–X (\(X =\) Mn, Fe, Co, Zn) exchange interactions are found to evolve from antiferromagnetic to ferromagnetic upon substitution.

Furthermore, inclusion of SOC profoundly modifies the band topology of pristine \(\mathrm{EuMn_2Bi_2}\). SOC closes the band gap in the antiferromagnetic state and induces band touching near the Fermi level, which in turn produces a band inversion between the Bi $s$ and Bi $p_{x,z}$ orbitals along the $\Gamma-A$ direction, giving rise to topological features such as Dirac and Weyl fermions. Four symmetry-related Weyl points are observed near \(E_F\), appearing in \(\pm\mathbf{k}\) pairs with opposite chirality due to inversion symmetry, and they act as monopole-like sources and sinks of Berry curvature. The Berry curvature hotspots, gapless surface states, and surface-projected Fermi arcs provide clear evidence for the emergence of a magnetic Weyl semimetal phase in this system.

The coexistence of broken time-reversal symmetry and non-trivial band topology makes \(\mathrm{EuMn_2Bi_2}\) a promising candidate for topological spintronic applications. The Weyl fermions and their associated Berry curvature can enable efficient spin–charge interconversion, while the robust Fermi arc states can serve as scattering-immune spin-polarized transport channels. Overall, our findings highlight \(\mathrm{EuMn_2Bi_2}\) as a versatile platform for engineering correlated magnetic topological phases and developing next-generation spintronic devices.

\bibliographystyle{unsrtnat}
\bibliography{MX}

\end{document}